\documentstyle[aps,epsfig,multicol]{revtex}

\begin{document}
\title{Penetration Depth Measurements in MgB$_2$: Evidence 
for Unconventional Superconductivity}

\author{C. Panagopoulos$^{1}$, B.D. Rainford$^{2}$, T. Xiang$^{3}$, 
C.A. Scott$^{2}$, M. Kambara$^{1}$ and I.H. Inoue$^{1,4,}$\cite{isao}}

\address{$^{1}$ Cavendish Laboratory and Interdisciplinary Research 
Centre for Superconductivity, University of Cambridge, 
Cambridge CB3 0HE, United Kingdom}

\address{$^{2}$ Department of Physics and Astronomy, University of 
Southampton, Southampton S017 1BJ, United Kingdom}

\address{$^{3}$ Institute of Theoretical Physics, Academia Sinica, 
P.O. Box 2735, Beijing 100080, People's Republic of China}

\address{$^{4}$ Electrotechnical Laboratory, Tsukuba 305-8568, Japan.}

\date{\today}

\maketitle

\begin{abstract}
We have measured the magnetic penetration depth of the recently discovered binary
superconductor MgB$_2$ using muon spin rotation and low field $ac$-susceptibility. 
From the damping of the muon precession signal we find the
penetration depth at zero temperature is $\sim 85$nm.
The low temperature penetration depth shows a quadratic temperature
dependence, indicating the presence of nodes in the superconducting energy
gap.
\end{abstract}


\begin{multicols}{2}

The discovery of superconductivity in the simple binary compound MgB$_2$
with a remarkably high transition temperature $T_c\sim 39$K has attracted
great interest \cite{Nagamatsu}. MgB$_2$ is a hexagonal AlB$_2$-type
compound, consisting of alternating hexagonal Mg layers and graphite-like B
layers. To explore the mechanism of superconductivity in this material it is
important to determine the symmetry of the superconducting order parameter
which governs the behavior of quasiparticle excitations below $T_c$.
Experimentally this can be done by measuring thermodynamic responses of
superconducting quasiparticles at low temperatures. In conventional $s$-wave
superconductors, there are no quasiparticle excitations at low energies and
the \thinspace thermodynamic and transport coefficients decay exponentially
at low temperatures. However, in unconventional superconductors with gap
nodes, such as in high-$T_c$ oxides, power law behaviors are expected in
thermodynamic coefficients at low temperatures.

Recently, the scanning tunneling 
conductance\cite{Rubio,Karapetrov,Sharoni,Schmidt}, 
the nuclear spin-lattice relaxation
rate\cite{Kotegawa} of MgB$_2$, and the specific heat\cite{Walti} have been measured. 
Although it was claimed that the experimental data are consistent 
with a conventional $s$-wave pairing gap, the data reported are 
rather controversial and no
consensus can really be reached regarding the pairing symmetry. The gap
values obtained from the scanning tunneling measurements vary from 2meV to 
7meV. The low temperature dependences of tunneling spectra reported by
different experimental groups also behave differently. The difference in
both the value of the energy gap and the low temperature dependence of the
tunneling spectrum is probably due to the state of the
surface, as well as the inhomogeneity of the samples measured. 
The spin-lattice relaxation rate
measured by Kotegawa $et$ $al$\cite{Kotegawa} shows a very small coherence
peak just below $T_c$, followed by an exponential decay in a broad
temperature range. However, their measurement was done at relatively high
temperatures (above $15$K), and it is not known whether this exponential
behavior extends to lower temperatures.

In this paper, we report our experimental data of the magnetic
penetration depth $\lambda $ of MgB$_2$ in the superconducting state. 
We have measured $\lambda $ using the transverse-field muon spin
rotation (TF-$\mu $SR) and low field $ac$
-susceptibility. From both measurements, we find that
at low temperatures $\lambda$ varies quadratically with temperature and
does not show the activated exponential behavior expected for a conventional 
$s$-wave superconductor. This quadratic behavior of $\lambda$ 
indicates the existence of nodes in the
superconducting energy gap of MgB$_2$.

The penetration depth is inversely proportional to the square root of the 
superfluid density. Its temperature dependence is a measure of 
the low-lying superconducting quasiparticle excitations and no 
phonon contribution is directly involved. This makes the analysis of   
penetration depth data simple. Another advantage of 
the penetration depth measurement is that it 
allows us to determine whether there are nodes in the 
superconducting energy gap even with slightly impure 
samples\cite{Bonn,Hirschfeld}.

The sample measured was commercially available MgB$_2$ powder (Alfa Aesar).
The superconducting transition temperature, as determined by both 
$ac$-susceptibility and dc-SQUID
(at 20G) measurements, is 37.5K. High field dc-SQUID and electron microscopy 
investigations showed less than 1\% of impurities present. 
The $ac$-susceptibility measurements were performed at an applied 
field $H_{ac}$=1G $RMS$ and frequency $f$=333Hz on fine  
powder. The absence of weak links was confirmed by checking the 
linearity of the pick-up voltage at 4.2K for $H_{ac}$ from 1 to 10G $RMS$ 
and $f$ from 16 to 667Hz.

TF-$\mu $SR is a sensitive technique for measuring $\lambda $. In this
technique, the field distribution of a flux-line lattice of a type-II
superconductor produced by an applied magnetic field is probed by 
fully polarised positive muons implanted into a specimen.
The muon decays with a life time $2.2\mu$s, emitting a positron
preferentially in the direction of the muon spin at the time of decay. 
By accumulating time histograms of the decay
positrons the muon polarisation can be followed as a function of
time. In type-II superconductors, the muon
spin precesses about the local field which is modulated by flux vortices.
The time resolved polarisation signal is oscillatory with a decreasing
amplitude. The damping of the muon precession signal provides a
measure of the inhomogeneity of the magnetic field $\Delta $B in the 
vortex state, hence the magnetic penetration depth 
$\lambda$\cite{Pincus,Barford,Uemura}. For polycrystalline samples 
the envelope of the muon precession signal has approximately a Gaussian 
form $exp(-\sigma ^2t^2/2)$ and the depolarisation rate $\sigma $ can 
be shown to be proportional to the superfluid density 
$1/\lambda^2$\cite{Barford,Uemura}. 
For isotropic type II superconductors, $\lambda $ is given by\cite{Pincus}
\begin{equation}
\sigma (\mu s^{-1})=7.904\times 10^4\times \lambda ^{-2}(nm). 
\end{equation}
whereas for anisotropic superonductors such as the high-$T_c$ cuprates, 
the in-plane penetration depth can be determined from 
$\sigma$ \cite{Barford} 
\begin{equation}
\sigma (\mu s^{-1})=7.086\times 10^4\times \lambda^{-2}(nm).  \label{musr}
\end{equation}

Evidence about the anisotropy of the superconductivity may be
obtained from the form of the distribution of internal fields P(B), which
can be derived by Fourier transforming the muon precession signal.
Detailed information requires data from single crystals. However in
powder samples of anisotropic superconductors it is found that the
distribution P(B) has a characteristic shoulder on the low field side of
the central frequency \cite{Weber}. We used the maximum entropy method to
extract P(B) from the TF-$\mu$SR data for MgB$_2$. For temperatures just
below $T_c$ the low field shoulder is clearly evident, so we conclude that
MgB$_2$ is anisotropic.

Our TF-cooled muSR measurements were
performed at the ISIS Facility, Rutherford Appleton Laboratory. 
The sample measured by TF-$\mu$SR was a pellet of MgB$_2$, 
with 4cm in diameter and 2mm thick,  prepared by cold pressing the powder. 
The pellet was mounted in a transverse
magnetic field $H_{app}$ which was above the lower critical field but below
the upper critical field, $H_{c1}<H_{app}<H_{c2}$. $H_{c1}$ and $H_{c2}$ of
MgB$_2$ are about 300G and 18T, respectively\cite{Bud'ko}. We used a field of 450G. 
(Measurements at not too high fields compared to $H_{c2}$
can minimise possible effects of dissipation due to flux motion and allow us
to obtain reliable data for the temperature dependence of the relaxation rate
$\sigma $\cite{Panagopoulos}.) A set of measurements at different fields (up to 600G)
was also done to ensure that the values of $\sigma $ obtained were
independent of the applied field.

The low-field $ac$-susceptibility is also a commonly used technique for
measuring $\lambda $. It has been successfully applied to high-$T_c$ 
materials \cite{CPTX} and is particularly suitable for
powder samples. The accuracy in the temperature dependence of $\lambda $
determined from the $ac$-susceptibility technique is significantly higher 
than that of TF-$\mu$SR\cite{Panagopoulos}. 
In the superconducting state, 
the effective $ac$-susceptibility $\chi $ of a powder sample is related to $\lambda $ 
by the equation \cite{Shoenberg,Porch}
\begin{equation}
\chi (T)=\chi _0\left\langle 1-\frac{3\lambda }R\coth \frac R\lambda +\frac{
3\lambda ^2}{R^2}\right\rangle ,
\end{equation}
where $\chi _0$ is the susceptibility in an ideal diamagnetic system, $R$ is
the radius of a grain, and $\left\langle \cdots \right\rangle $ denotes a
grain average defined by $\left\langle x\right\rangle \equiv \int
dRxR^3g(R)/\int dRR^3g(R)$ with $g(R)$ the grain size distribution function.
At low temperatures, $\chi (T)$ can be expanded with $\delta \lambda
(T)=\lambda (T)-\lambda _0$, where $\lambda _0=\lambda (0K)$. To the leading
order in $\delta \lambda $, we find
\begin{equation}
\chi (T)\approx \chi (0)+\alpha \delta \lambda (T), \label{sus}
\end{equation}
where 
\begin{eqnarray}
\chi (0) &=&\chi _0\left\langle 1-\frac{3\lambda _0}R\coth \frac R{\lambda _0
}+\frac{3\lambda _0^2}{R^2}\right\rangle , \\
\alpha &=&\chi _0\left\langle \frac{6\lambda _0}R-3\coth \frac R{\lambda _0}-
\frac{3R}{\lambda _0\sinh ^2[R/\lambda _0]}\right\rangle <0.
\end{eqnarray}
At low temperatures, the change in $\lambda $ is proportional to the change
in $\chi $. Thus from the temperature dependence of $\chi (T)$ 
(exponential or power law),  one can readily determine the temperature
dependence of $\lambda (T)$ in the low temperature regime. 
However, to determine the absolute value of $\lambda$, we need to know 
accurately the grain distribution function $g(R)$ which is beyond the 
scope of the present work.

\begin{figure}[htb]
\begin{center}
\epsfig{file=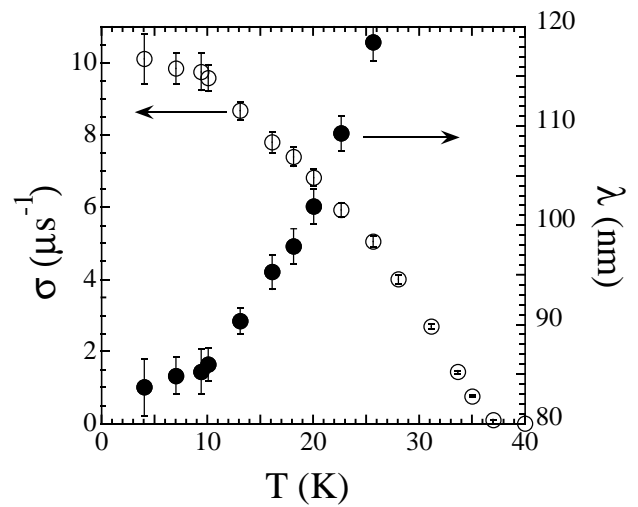,clip=,width=85mm,angle=0}
\end{center}
\caption{The muon depolarisation rate $\sigma$ for MgB$_2$ and the
corresponding penetration depth $\lambda$ as determined from Eq. (\ref
{musr}) versus temperature.}
\label{fig1}
\end{figure}

Fig. 1 shows the temperature dependence of the muon
depolarisation rate $\sigma $. At zero temperature we find that 
$\sigma (0K)\sim 10\mu s^{-1}$. The corresponding in-plane penetration 
depth at 0K is $\lambda_0\sim$85nm. The temperature dependence of 
$\lambda $ at low temperature is clearly
stronger than the exponentially activated temperature dependence expected 
for an $s$-wave superconductor. In fact, 
we find that $\sigma $ varies approximately 
quadratically with $T$ in the whole temperature range 
below $T_c$ (Fig. 2). This is supported by our low temperature $ac$-susceptibility
data shown in Fig. 3, which is related to the penetration depth by Eq. (\ref{sus}). 
This $T^2$ behavior of $\lambda$ at low temperatures shows unambiguously
that there are nodes in the superconducting energy gap of MgB$_2$\cite{Hirschfeld}. 

\begin{figure}[htb]
\begin{center}
\epsfig{file=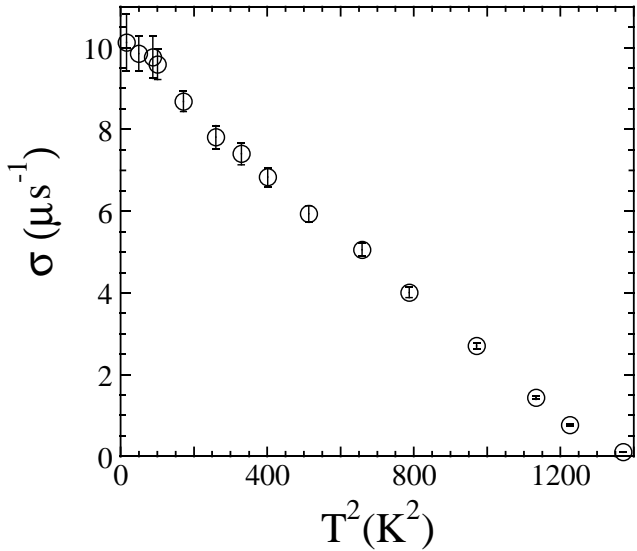,clip=,width=85mm,angle=0}
\end{center}
\caption{The muon depolarisation rate $\sigma$ versus $T^2$.}
\label{fig2}
\end{figure}

\begin{figure}[htb]
\begin{center}
\epsfig{file=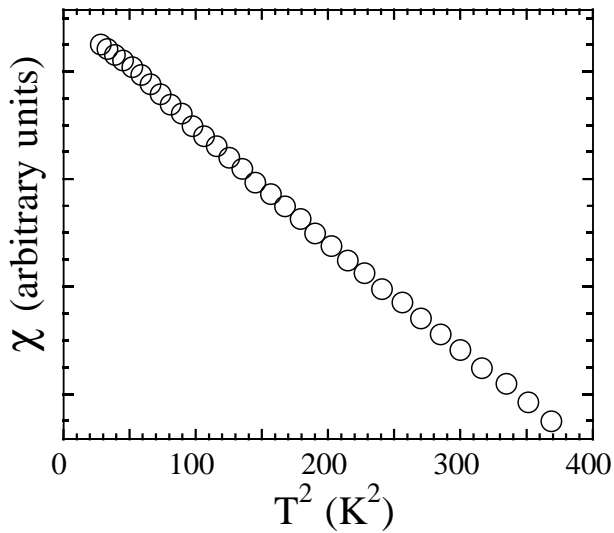,clip=,width=85mm,angle=0}
\end{center}
\caption{The $ac$-susceptibility $\chi$ as a function of $T^2$
at low temperatures. The measurement was performed at 1G and 
333Hz.}
\label{fig3}
\end{figure}

The $\lambda $ values shown in Fig. 1 have been estimated using 
Eq. (\ref{musr}). If the electromagnetic response of MgB$_2$ is 
more isotropic, the values of $\lambda $ reported here need to 
be multipied by a constant factor $c$ of order 1. In the extremely 
isotropic case $c=1.06$. However, the $T^2$ dependence 
of $\lambda$ is unchanged.

The $T^2$ behavior of $\lambda $ at low temperatures is a typical feature of
disordered superconductors with line nodes, such as the Zn-doped high-$T_c$
superconductor YBa$_2$Cu$_3$O$_7$ \cite{Bonn,PanagopoulosZn}. 
In fact as little as 0.31 percent Zn substitution can cause a 
crossover from a  linear temperature dependence to $T^2$ without 
affecting $T_c$\cite{Bonn}. This has been interpreted as the effect 
of impurity scattering on a superconductor with line nodes in the energy 
gap and offers a natural explanation for the $T^2$ behavior of $\lambda $ 
we found in MgB$_2$. The $T^2$ dependence of $\lambda$ 
in disordered high-$T_c$ oxides is robust
against Zn doping\cite{PanagopoulosZn}. 
In MgB$_2$ it has also been found that this $T^2$ behavior of 
$\lambda $ is robust against non-magnetic (Zn and Ca) 
doping\cite{McAllister}.

In conclusion, we have measured the magnetic penetration depth of the newly
discovered superconductor MgB$_2$ using the transverse-field $\mu $SR and
low-field ac-susceptibility techniques. The value of $\lambda$ at 0K 
is about 85nm. $\lambda $ shows a $T^2$ dependence
at low temperatures. This is strong evidence for unconventional
superconducting pairing in MgB$_2$. 

We are grateful to A.D. Taylor of the ISIS Facility, Rutherford Appleton
Laboratory for the allocation of muon beam time. We thank J.R. Cooper for
useful discussions and Y. Shi for help in the characterisation
measurements. C.P. thanks the Royal Society for financial support. T.X.
acknowledges the hospitality of the Interdisciplinary Research Center in
Superconductivity of the University of Cambridge, where this work was done,
and the financial support from the National Natural Science Foundation of
China.

\end{multicols}


\begin{references}
\bibitem[\dagger]{isao}  Present address: Correlated Electron Research
Center (CERC), Tsukuba 305-8562, Japan.

\bibitem{Nagamatsu}  J. Nagamatsu, N. Nakagawa, T. Muranaka, Y. Zenitani and
J. Akimitsu, Nature (London), {\bf 410}, 63 (2001).

\bibitem{Rubio}  G. Rubio-Bollinger, H. Suderow, and S. Vieira,
cond-mat/0102242.

\bibitem{Karapetrov}  G. Karapetrov, M. Iavarone, W. K. Kwok, G. W. Crabtree
and D. G. Hinks, cond-mat/0102312.

\bibitem{Sharoni}  A. Sharoni, I. Felner, and O. Millo Racah,
cond-mat/0102325.

\bibitem{Schmidt}  H. Schmidt, J. F. Zasadzinski, K. E. Gray, and D. G.
Hinks, cond-mat/0102389.

\bibitem{Kotegawa}  H. Kotegawa, K. Ishida, Y. Kitaoka, T. Muranaka, and J.
Akimitsu, cond-mat/0102334.

\bibitem{Walti} Ch. Walti, E. Felder, C. Degen, G. Wigger, R. Monnier, b. Delley, 
and H. R. Ott, cond-mat/0102522. 

\bibitem{Bonn}  D.A. Bonn, S. Kamal, K. Zhang, R. Liang, B.J. Baar, 
E. Klein, and W.N. Hardy,  Phys. Rev. B {\bf 50}, 4051 (1994).

\bibitem{Hirschfeld}  P.J. Hirschfeld and N. Goldenfeld, Phys. Rev. 
B {\bf48}, 4219 (1993). 

\bibitem{Pincus}  P. Pincus, A.C. Gossard, V. Jaccarino and J.H. Wernick, 
Phys. Lett. {\bf 13}, 21 (1964).

\bibitem{Barford}  W. Barford and J.M.F. Gunn, Physica (Amsterdam) 
{\bf 156}, 515 (1987).

\bibitem{Uemura}  Y.J. Uemura {\it et al.}, Phys. Rev. B {\bf 38}, 909 (1988); 
C. Bernhard {\it et al.}, Phys. Rev. B {\bf 52}, 10488 (1995); 
B. Pumpin {\it et al.}, Phys. Rev. B {\bf 42}, 8019 (1990); 
J. E. Sonier {\it et al.}, Phys. Rev. Lett. 
{\bf 72}, 744 (1994). 

\bibitem{Weber} M.Weber, A.Amato, F.N.Gygax, A.Schenck, H.Maletta,
V.N. Duginov, V.G.Grebinnik, A.B.Lazarev, V.G.Olshevsky, V.Y.Pomjakusin, 
S.N.Shilov, V.A.Zhukov, B.F.Kirillov, A.V.Pirogov, A.N.Ponomarev, V.G.Storchak,
S.Kaputsa and J.Bock, Phys. Rev. B {\bf 48}, 13022 (1993).

\bibitem{Bud'ko}  S. L. Bud'ko, C. Petrovic, G. Lapertot, C. E. Cunningham,
P. C. Canfield, M-H. Jung, A. H. Lacerda, cond-mat/0102413.

\bibitem{Panagopoulos}  C. Panagopoulos, B.D. Rainford, J.R. Cooper, W. Lo,
J.L. Tallon, J.W. Loram, J. Betouras, Y.S. Wang and C.W. Chu, Phys. Rev. B 
{\bf 60}, 14617 (1999).

\bibitem{CPTX}  C. Panagopoulos and T. Xiang, Phys. Rev. Lett. {\bf 81},
2336 (1998).

\bibitem{Shoenberg}  D. Shoenberg, Superconductivity (Cambridge University
Pres, Cambridge, 1954) p.164.

\bibitem{Porch}  A. Porch, J.R. Cooper, D.N. Zheng, J.R. Waldram, A.M.
Campbell and P.A. Freeman, Physica (Amsterdam) C {\bf 214}, 350 (1993).


\bibitem{PanagopoulosZn}  C. Panagopoulos, J.R. Cooper, N. 
Athanassopoulou and J. Chrosch, Phys. Rev. B {\bf54}, R12721 (1996).

\bibitem{McAllister}  J.A. McAllister (private communication). 

\end{references}
\end{document}